\documentclass[twoside]{ilcws08}
\usepackage[latin1]{inputenc}
\usepackage[dvips]{graphicx,epsfig,color}
\usepackage{wrapfig,rotating}
\usepackage{amssymb,amsmath,array}

\begin{document}
\title{
Kinematic Fitting in the Presence of ISR at the ILC} 
\author{Jenny List$^1$, Moritz Beckmann$^{1,2}$ and Benno List$^3$
\vspace{.3cm}\\
1 - DESY Hamburg, Germany
\vspace{.1cm}\\
2 - Leibniz Universit\"at Hannover, Germany
\vspace{.1cm}\\
3 - Universit\"at Hamburg, Germany\\
}

\maketitle
\begin{abstract}
Kinematic fitting is a well-established tool to improve jet energy and invariant mass resolutions by fitting the measured values under constraints (e.g. energy conservation).\\
However, in the presence of substantial ISR and Beamstrahlung, na\"ive energy and (longitudinal) momentum constraints fail due to the a priori unknown amount of undetected momentum carried away by collinear photons. It is possible to take care of those two effects and thus obtain significantly higher mass resolutions.
\end{abstract}

\section{Introduction}
Kinematic fitting is used to improve measured quantities and/or determine unmeasured quantities by using known constraints of the events, e.g. momentum conservation. Already at LEP this led to a significant improvement of the W mass resolution ~\cite{lep}. The special new problems at the ILC are the  substantially higher amounts of initial state radiation (ISR) and Beamstrahlung which are due to the increase in beam energy and luminosity.

\newcommand{\Eslash}  {E\kern-0.6em\slash}

\section{Benchmark Sample}
\label{subsec:sample}
\begin{wraptable}{r}{0.45\columnwidth}
	\vspace*{-7mm}
	\centerline{\begin{tabular}{l|rr}
		Cut & events & \% \\
		\hline
		total sample                        & 52\,490 & 100.0\% \\
		$|\cos(\theta_\text{jet})| < 0.989$ & 39\,940 &  76.1\% \\
		$N_\text{track}/\text{jet} \geq 1$  & 38\,925 &  74.2\% \\
		$E_\text{jet} > 4.5$ GeV            & 38\,912 &  74.1\% \\
	\end{tabular}}
	\caption{Cuts applied to jets.}
	\label{tab:cuts}
	\vspace*{-4mm}
\end{wraptable}

To show the problem and to test possible remedies, a sample $e^+ e^- \rightarrow u\bar{d}d\bar{u}$ with an integrated luminosity of 15~fb$^{-1}$ is used. The quarks stem mostly from $W^+ W^-$ and about 1\% from $ZZ$ decays. A sample of light quarks has been chosen to ensure absence of B-mesons and subsequent leptonic decays. The sample is a full Monte-Carlo simulation of the LDCPrime\_02Sc detector model with full reconstruction using Pandora particle flow \cite{pandora}. The reconstructed particles are forced into four jets using the Durham Jetfinder \cite{durham}. In order to select well-reconstructed events, the cuts listed in Table~\ref{tab:cuts} are applied to these jets. Based on MC information, two subsamples are then selected to separately investigate events with and without a significant amount of missing energy $E_\text{miss}=\sqrt{s}-\sum E_\text{quark}$, where $\sqrt{s}$ is the nominal center of mass energy (500 GeV):
\begin{itemize}
	\item 'no \Eslash': $E_\text{miss} < 5$ GeV; 15\,726 events $\widehat{=} \, 40\%$ of sample after first cuts
	\item '\Eslash': $E_\text{miss} > 30$ GeV; $\cos\theta_\gamma > 0.999$, if $E_\gamma \geq 5$ GeV; 9\,186 events $\widehat{=} \, 24\%$
\end{itemize}
The remaining events (little missing energy or photons within detector acceptance) will be investigated later.
Five constraints are applied to both subsamples:
\begin{itemize}
	\item[] \#1-4: four-momentum conservation: $\sum p_\text{jets} = (\sqrt{s},0,0,0)$
	\item[] \#5: equal invariant di-jet masses for those jet pairs from $W_1/Z_1$ and $W_2/Z_2$
\end{itemize}
Since the pairing of the jets is a priori unknown, all three possible combinations are fitted. The ``best jet combination'' is the one with the highest fit probability $> 0$ and a di-jet mass of $50 \text{ GeV} < m_\text{jj} < 110 \text{ GeV}$.

\section{4-Jet Hypothesis used as Benchmark}
\begin{figure}[h]
   \setlength{\unitlength}{1.0cm}
   \begin{picture}(14.0, 4.7)
     \put( 0.0, -0.1)  {\epsfig{file=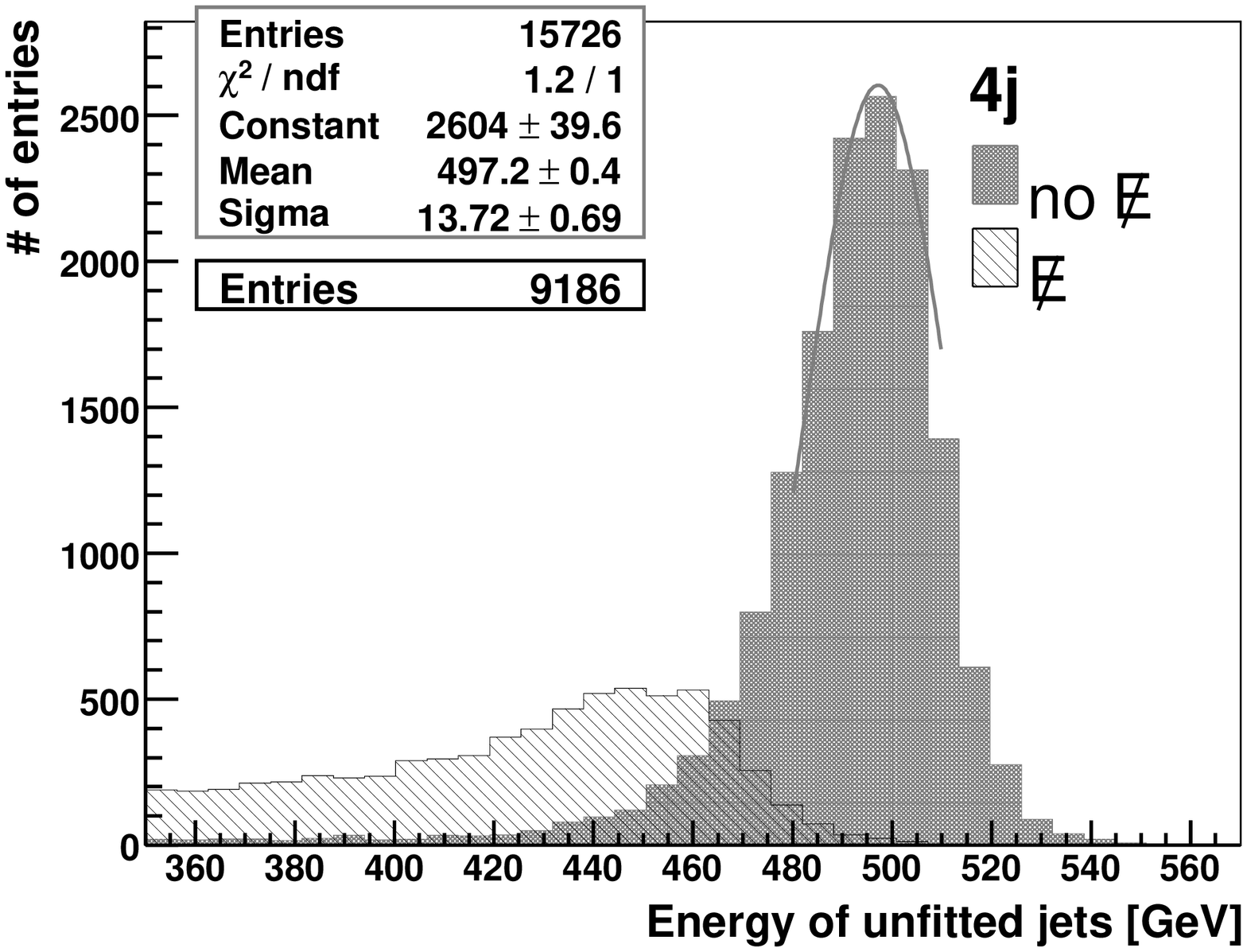, bb= 0 0 567 500, clip= , width=0.5\linewidth}}
     \put( 7.1, -0.1)  {\epsfig{file=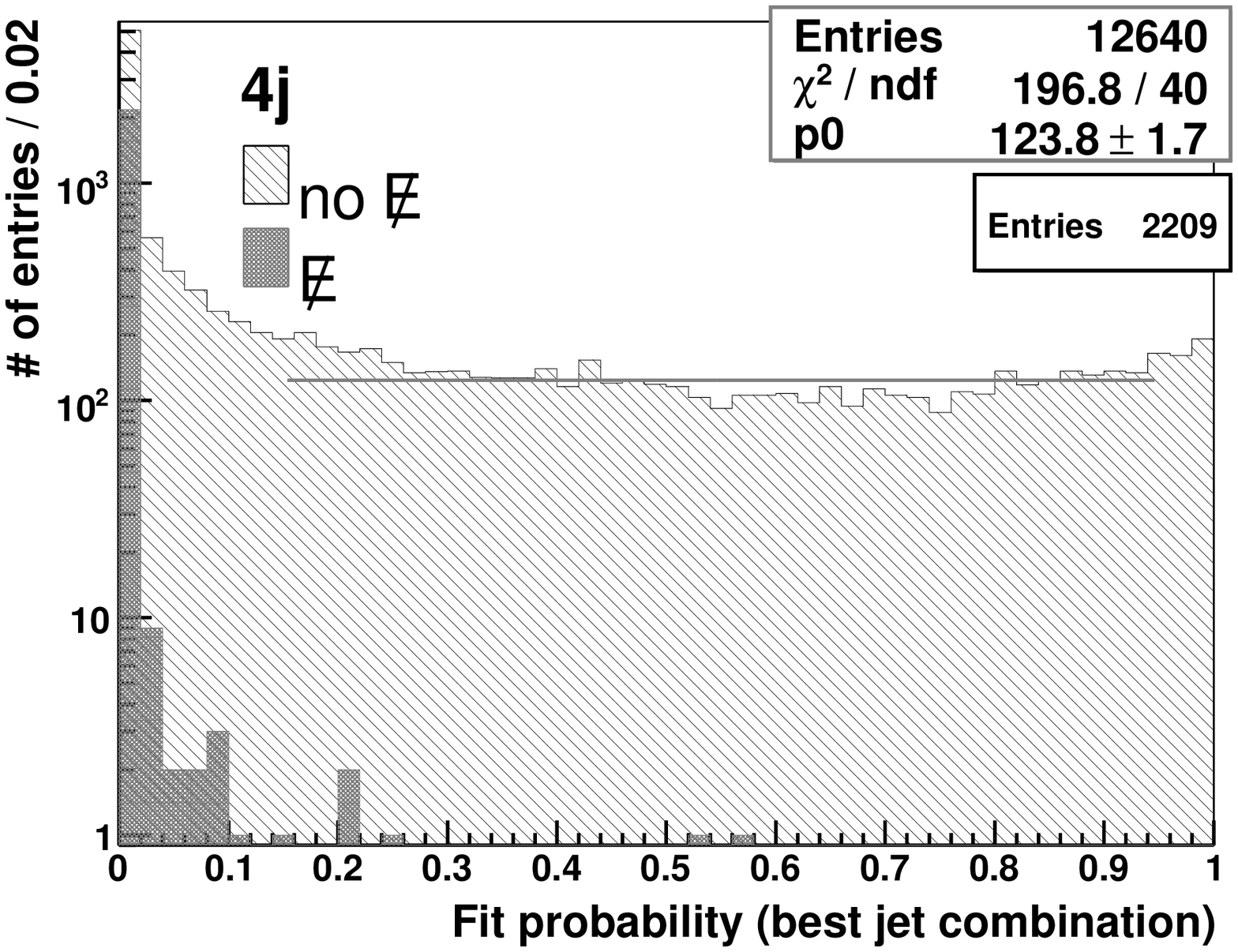, bb= 0 0 567 500, clip= , width=0.5\linewidth}}
     \put( 0.0, 0.5)  {(a)}
     \put( 7.1, 0.5)  {(b)}
   \end{picture}
   \vspace*{-7mm}
   \caption{Both subsamples: (a) sum of jet energies, (b) fit probability.}
   \label{Fig:DiffIsr}
\end{figure}
Fig.~\ref{Fig:DiffIsr}(a) shows the total visible energy for both subsamples.
In the case 'no \Eslash', the constraint $\sum E_j = 500$ GeV can be fulfilled within some energy spread (80\%), while a fit enforcing such a constraint has very little chance to converge in the other case (24\%). This can be seen in Fig.~\ref{Fig:DiffIsr}(b), where the fit probability for the best jet combination is shown (converged fits only).

\section{Modelling ISR and Beamstrahlung}
In this first approach, possible ISR and Beamstrahlung photons are modelled as one object with measured parameters and errors reflecting the missing momentum spectrum. Contrary to other approaches (unmeasured parameters, soft constraints), no hard constraints are lost.

The photons are parametrized by $p_x, p_y, p_z$. Since both ISR and Beamstrahlung are mostly emitted parallel to the beam, $p_x$ and $p_y$ are fixed to zero and only $p_z$ is varied by the fit. The ``measured'' value of $p_z$ is set to zero and the error is assumed to be Gaussian with $\sigma = 100$ GeV (first try, since easy to implement).

The MC simulation generates one photon each for both $e^+$ and $e^-$. However, in most cases only one of these two photons carries away a substantial amount of energy. Thus it is considered sufficient to add only one photon to the fit hypothesis as shown in the following.

\section{4-Jet-Plus-Photon vs. 4-Jet Hypothesis}
\subsection{'\Eslash' Subsample}
\begin{figure}[h]
   \setlength{\unitlength}{1.0cm}
   \begin{picture}(14.0, 9.8)
     \put( 0.0,  5.2)  {\epsfig{file=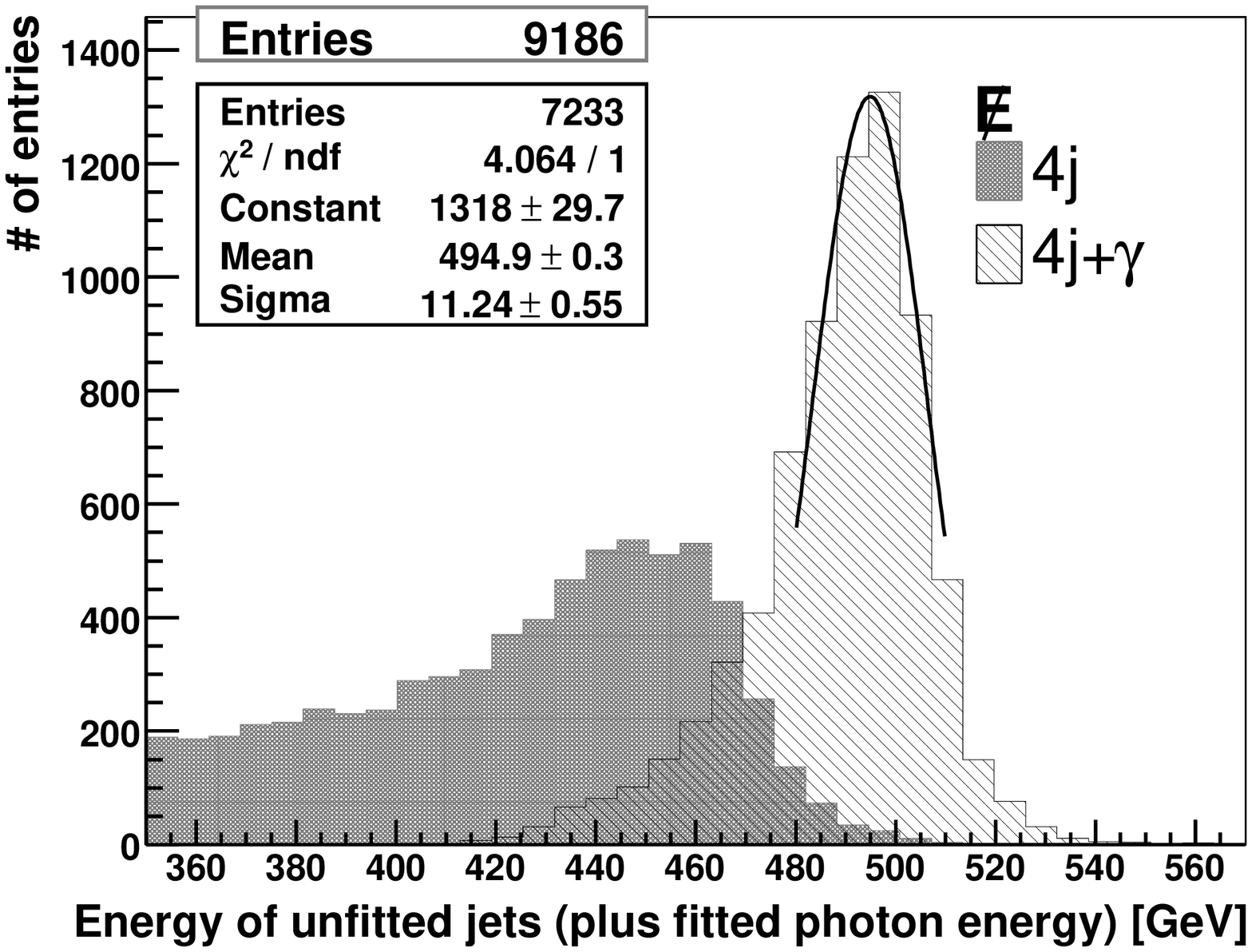, bb= 0 0 567 500, clip= , width=0.5\linewidth}}
     \put( 7.1,  5.2)  {\epsfig{file=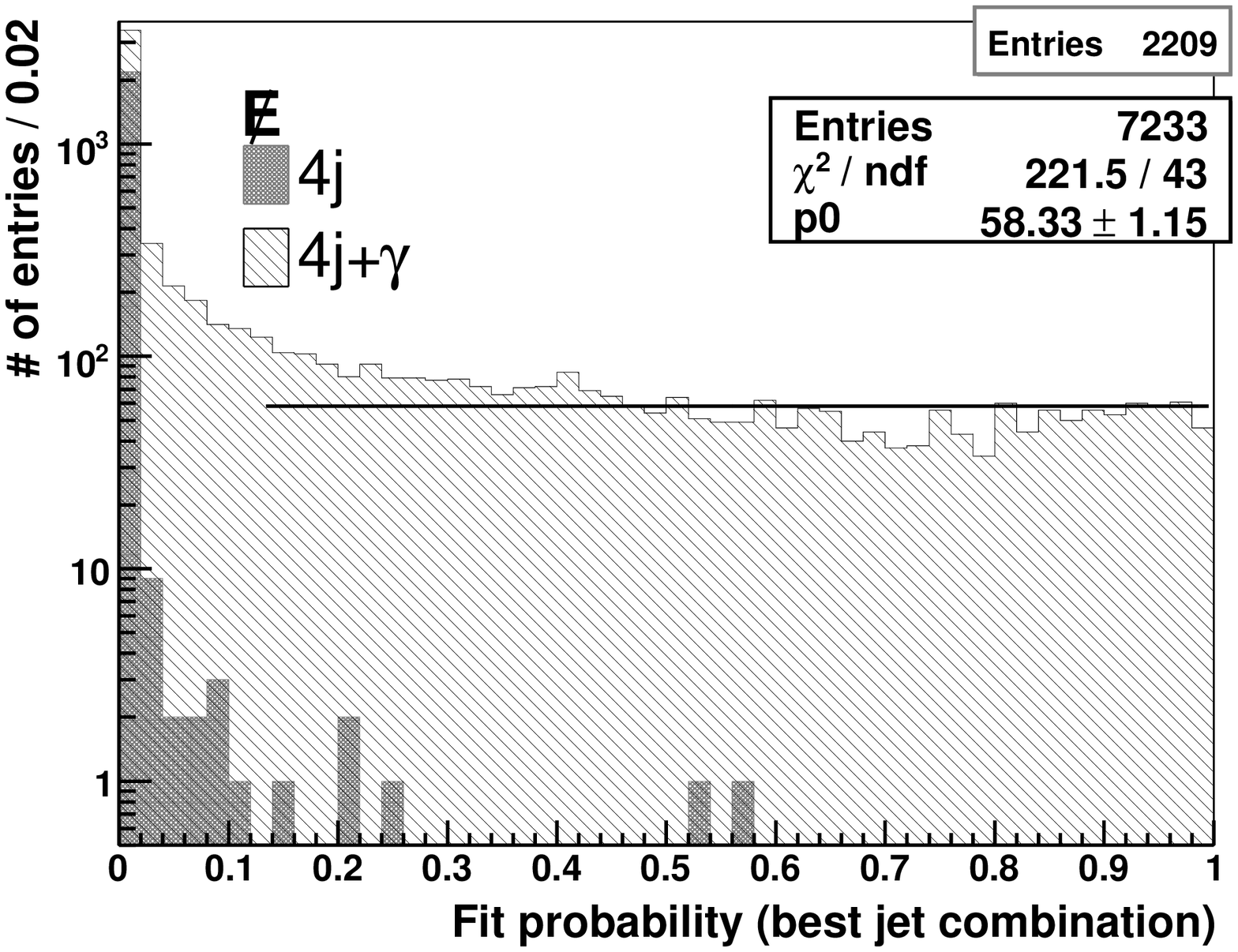, bb= 0 0 567 500, clip= , width=0.5\linewidth}}
     \put( 0.0, -0.1)  {\epsfig{file=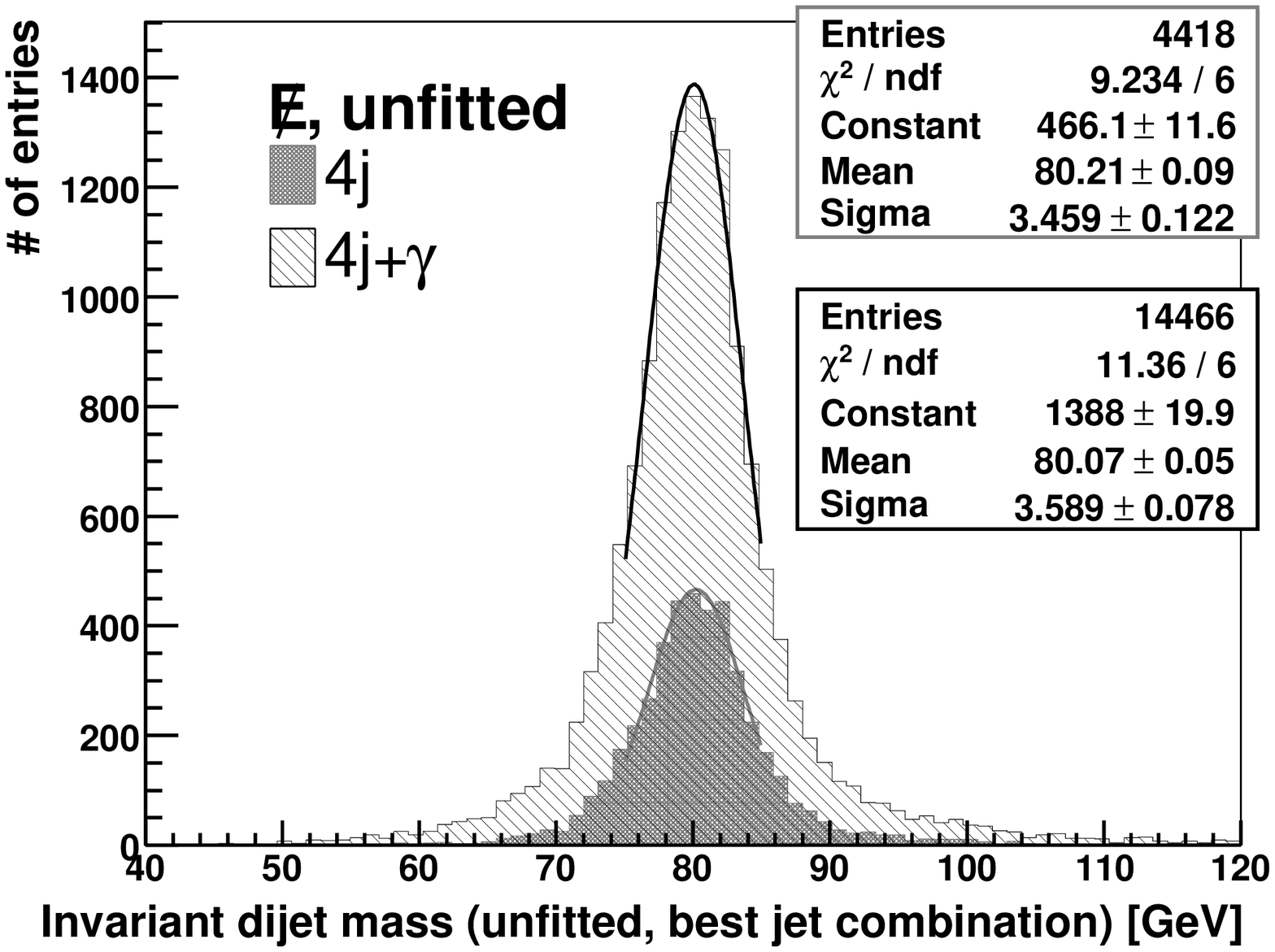, bb= 0 0 567 500, clip= , width=0.5\linewidth}}
     \put( 7.1, -0.1)  {\epsfig{file=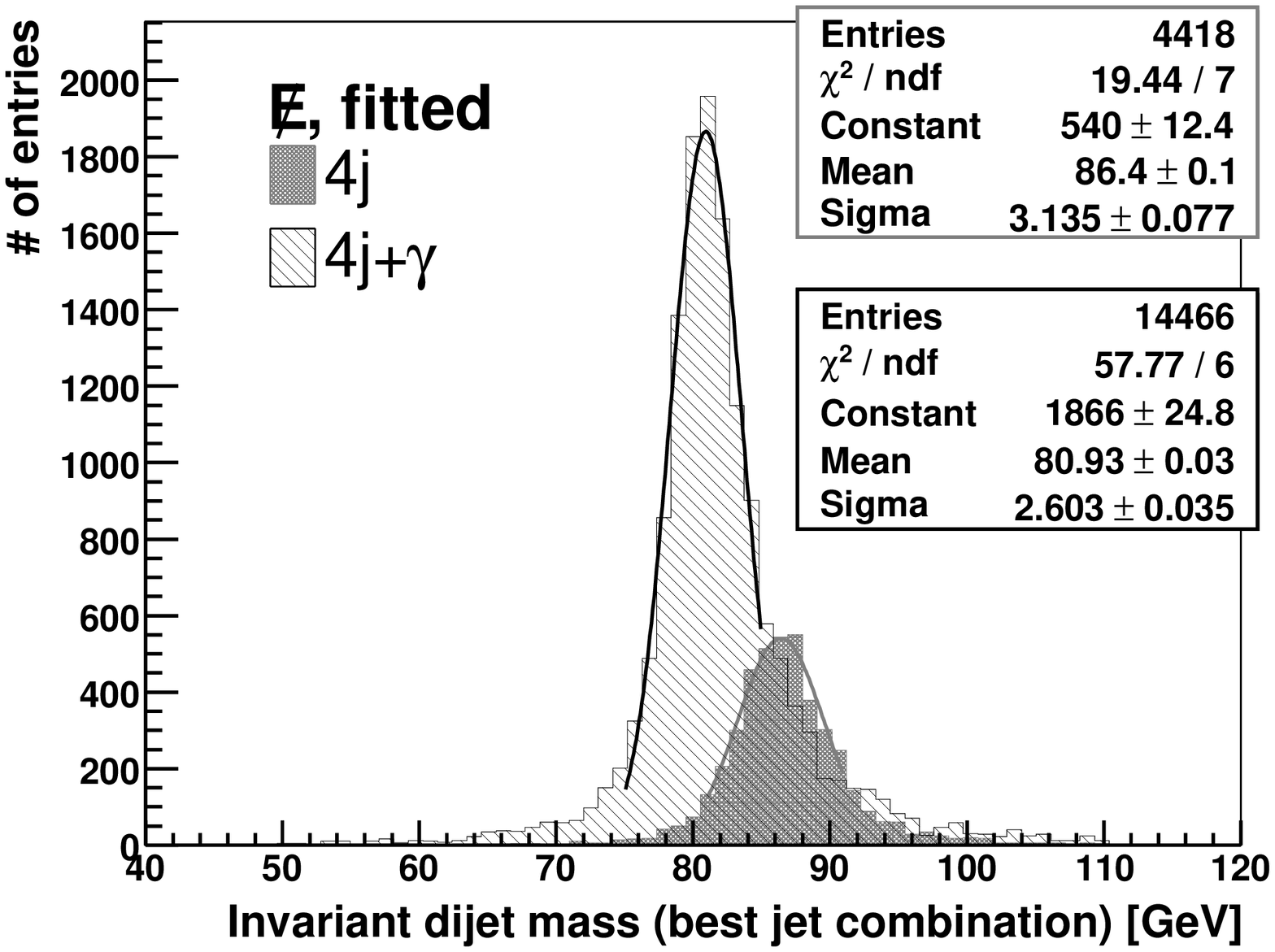, bb= 0 0 567 500, clip= , width=0.5\linewidth}}
     \put( 0.0, 5.8)  {(a)}
     \put( 7.1, 5.8)  {(b)}
     \put( 0.0, 0.5)  {(c)}
     \put( 7.1, 0.5)  {(d)}
   \end{picture}
   \vspace*{-7mm}
   \caption{Both fit hypotheses, tested on the '\Eslash' subsample: (a) sum of jet energies, (b) fit probability, (c) invariant di-jet masses before and (d) after the fit.}
   \label{Fig:PlotsIsr}
   \vspace*{-2.5mm}
\end{figure}
Fig.~\ref{Fig:PlotsIsr} shows a comparison of the two different hypotheses on the '\Eslash' subsample. Fig.~\ref{Fig:PlotsIsr}(a) shows again the total visible energy, for 4-jet+$\gamma$ the fitted photon energy is added. Now it is also possible for the fit to converge on the '\Eslash' subsample (79\%).
The fit probability distribution is shown in Fig.~\ref{Fig:PlotsIsr}(b).

\begin{wrapfigure}{l}{0.48\columnwidth}
   \vspace*{-5mm}
	\centerline{\includegraphics[width=0.46\columnwidth]{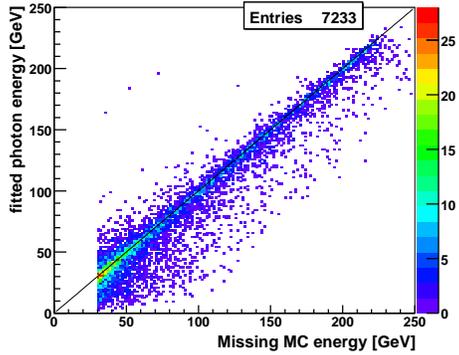}}
   \vspace*{-5mm}
	\caption{Missing energy in '\Eslash'  subsample: fitted vs. MC.}\label{Fig:PhotonEnergyIsr}
   \vspace*{-2mm}
\end{wrapfigure}
Fig.~\ref{Fig:PlotsIsr}(c) and (d) show the invariant 2-jet masses before and after the fit. In the 4-jet hypothesis the peak of the distribution is shifted towards higher masses, since the fit tries to compensate the missing energy to satisfy energy conservation. Thus, too large invariant masses are fitted, which makes the distinction between W and Z more difficult. The 4-jet+$\gamma$ fit includes the effect from ISR and Beamstrahlung and diminishes the bias and the mass peak width significantly. From the correlation between fitted photon energy and the missing MC energy $E_\text{miss}$ in Fig.~\ref{Fig:PhotonEnergyIsr} can be seen that the fit indeed recovers the energy of the escaping photons.
\subsection{'No \Eslash' Subsample}
\begin{figure}[h]
   \setlength{\unitlength}{1.0cm}
   \begin{picture}(14.0, 9.8)
     \put( 0.0,  5.2)  {\epsfig{file=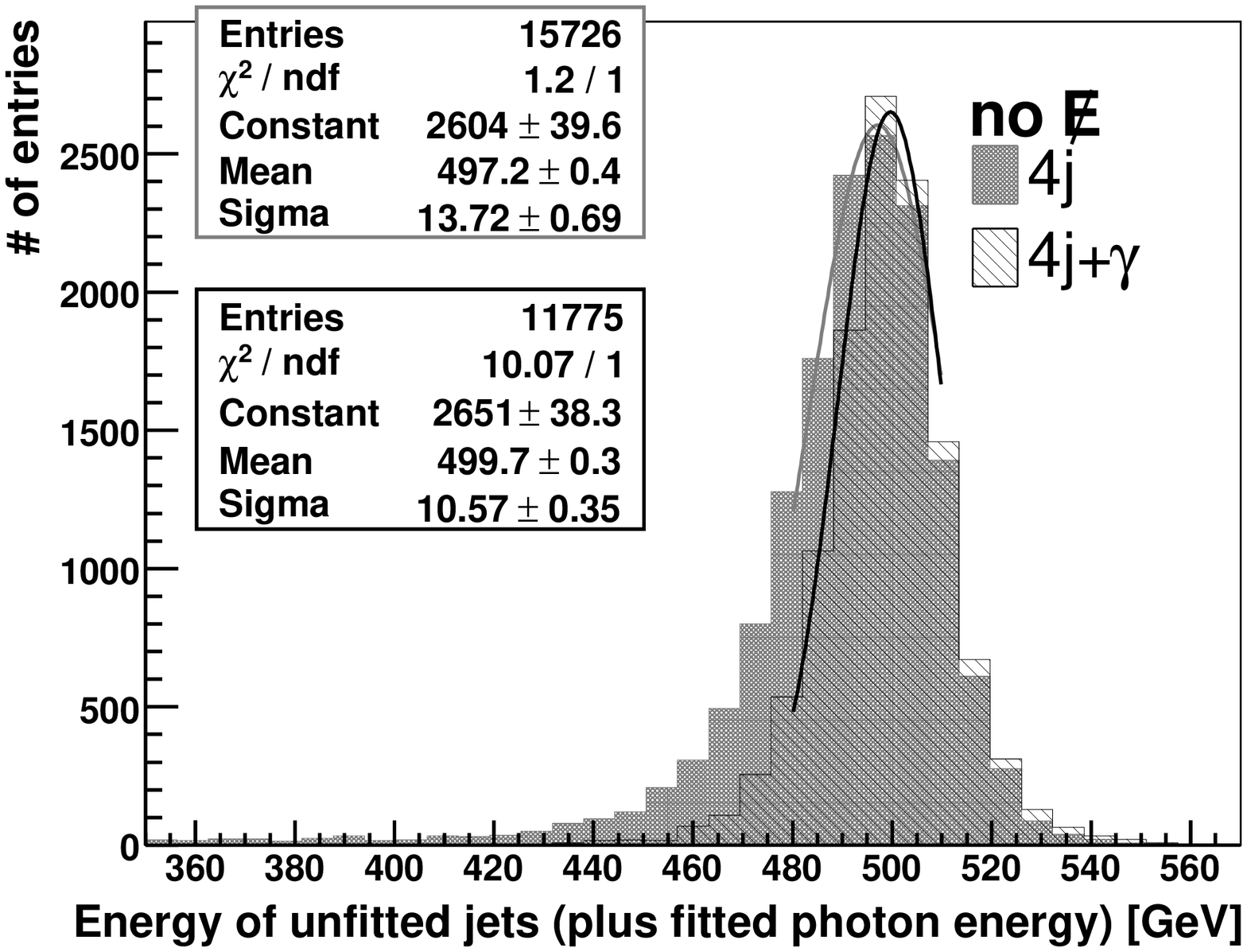, bb= 0 0 567 500, clip= , width=0.5\linewidth}}
     \put( 7.1,  5.2)  {\epsfig{file=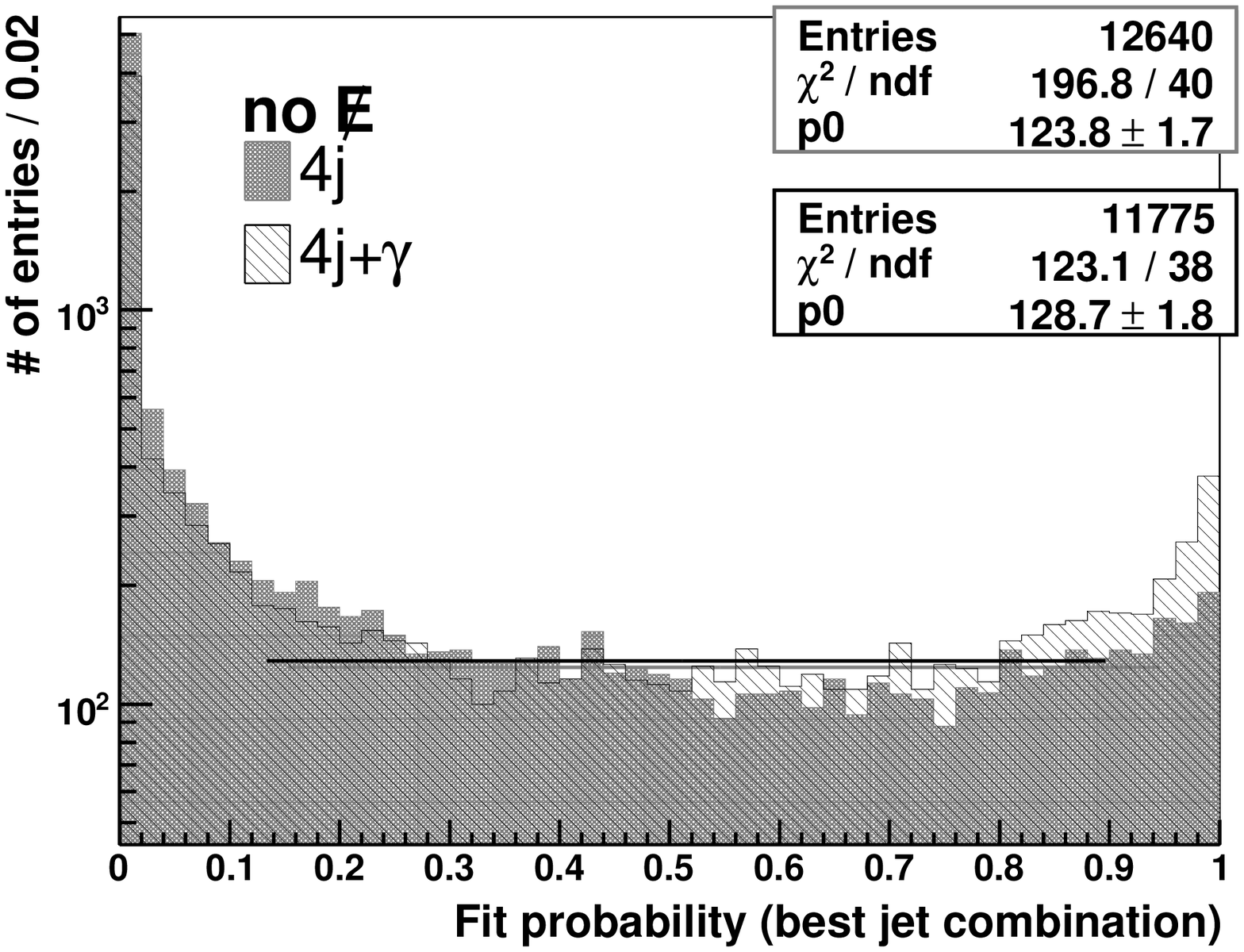, bb= 0 0 567 500, clip= , width=0.5\linewidth}}
     \put( 0.0, -0.1)  {\epsfig{file=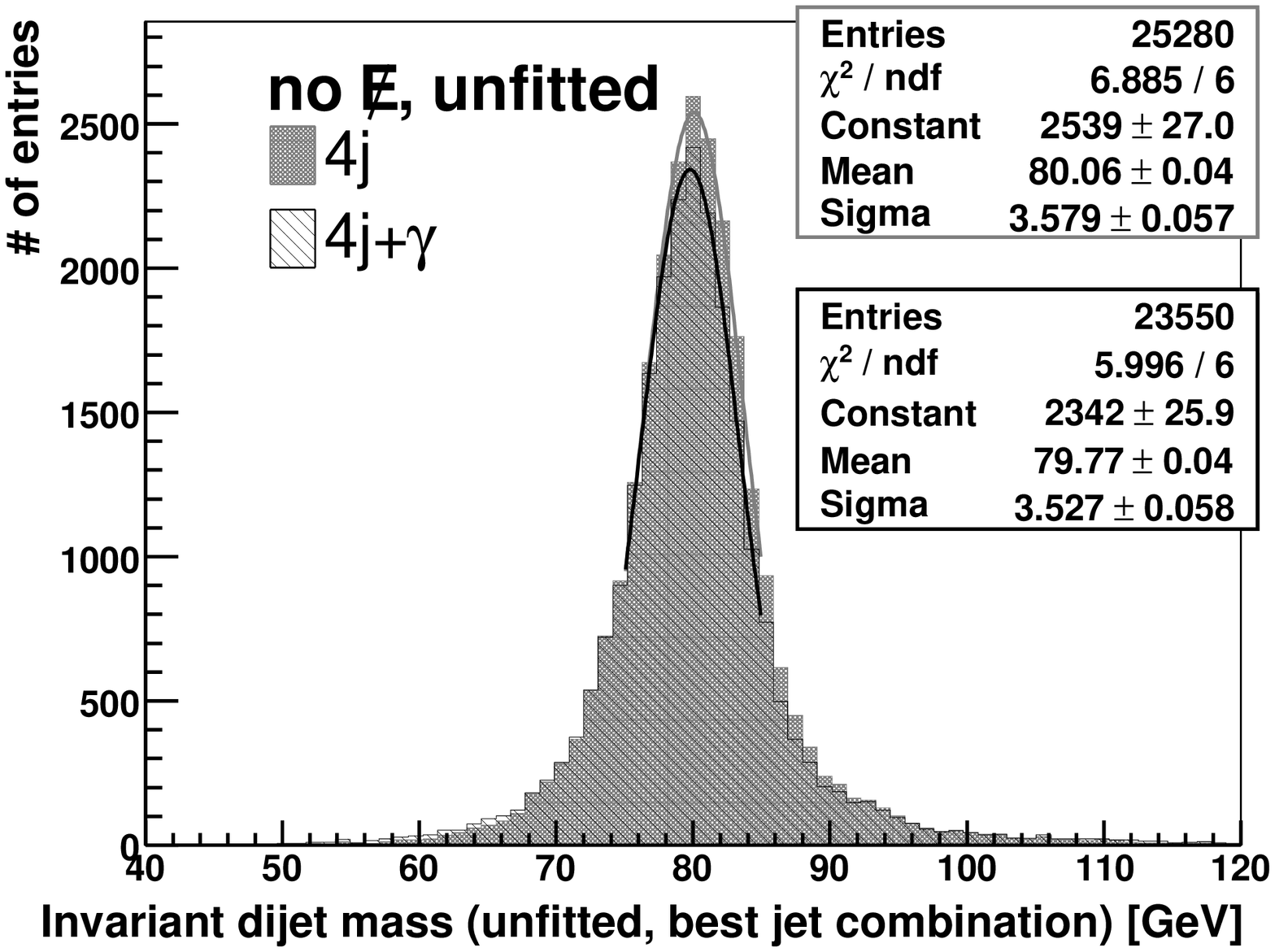, bb= 0 0 567 500, clip= , width=0.5\linewidth}}
     \put( 7.1, -0.1)  {\epsfig{file=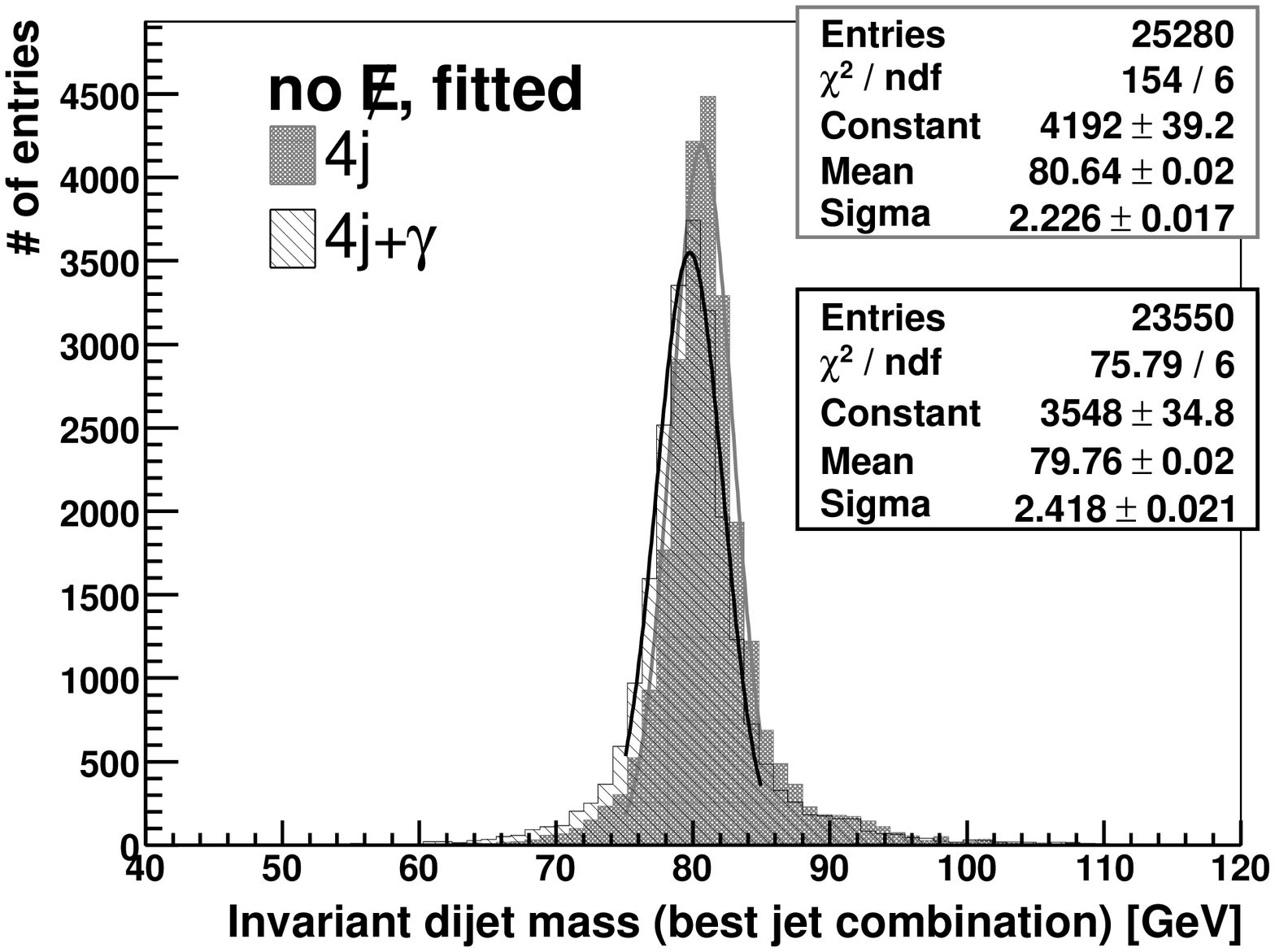, bb= 0 0 567 500, clip= , width=0.5\linewidth}}
     \put( 0.0, 5.8)  {(a)}
     \put( 7.1, 5.8)  {(b)}
     \put( 0.0, 0.5)  {(c)}
     \put( 7.1, 0.5)  {(d)}
   \end{picture}
   \vspace*{-7mm}
   \caption{Both fit hypotheses, tested on the 'no \Eslash' subsample: (a) sum of jet energies, (b) fit probability, (c) invariant di-jet masses before and (d) after the fit.}
   \label{Fig:PlotsNoIsr}
\end{figure}
Since, ideally, one fit should be sufficient for all events, the effects of the 4-jet+$\gamma$ hypothesis on the 'no \Eslash' subsample are investigated as well. Fig.~\ref{Fig:PlotsNoIsr} contains the distributions corresponding to Fig.~\ref{Fig:PlotsIsr} for this subsample. Fig.~\ref{Fig:PlotsNoIsr}(a) shows that small amounts of energy are recovered by the 4-jet+$\gamma$ fit ('no \Eslash' contains missing energy up to 5 GeV), but the fit converges for fewer events as shown in Fig.~\ref{Fig:PlotsNoIsr}(b) (to be investigated). Fig.~\ref{Fig:PlotsNoIsr}(c,d) illustrate that 4-jet+$\gamma$ yields slightly less resolution improvement, but eliminates the bias of the 4-jet fit.
\section{Summary and Outlook}
As the first results show, it is possible to reconstruct ISR photons and improve the di-jet mass resolution by about 1 GeV already with a very basic photon parametrization. Further improvement, especially concerning the convergence, can be achieved by a more accurate modelling of ISR and Beamstrahlung, better error estimation and jet energy scaling. The concept will be tested with other quark flavors and compared to a soft constraint approach. A special treatment for ISR photons measured in the detector has to be implemented.

%
%
%
%
%
%
\begin{footnotesize}

%
%
%
%
\end{footnotesize}


\end{document}